# Improving Anonymity in Shared Key Primitives Based on Perfect Hash Families

Mausumi Bose  and  Rahul Mukerjee
Applied Statistics Unit       Indian Institute of Management Calcutta
Indian Statistical Institute  Joka, Diamond Harbour Road
203 B T Road                  Kolkata 700 104, India
Kolkata 700 108, India

*Abstract*: We propose a new scheme for sharing symmetric key operations among a set of participants according to a $(t,n)$ threshold access structure. We focus on anonymity properties of this scheme and show that this scheme provides improved values of anonymity measures than the existing ones. In particular, the scheme can provide optimal and equitable participant anonymity when it is based on balanced perfect hash families.

## 1. Introduction

Schemes for sharing symmetric key operations such as block ciphers or message authentication codes have been of considerable recent interest, as evident from the significant contributions by Brickell et al. (2000), Martin et al. (2003, 2005), Long et al. (2006), Martin and Ng (2007), Zaverucha and Stinson (2010) and others. Threshold access structures play a crucial role in this context. A $(t,n)$ *threshold access structure* involves $n$ participants such that every group of $t$ participants has access to at least one key, while no set of $t-1$ or fewer participants has access to any key. Let $\Gamma$ be the collection of all groups of $t$ participants, and let $A$ denote any typical member of $\Gamma$. Thus, every $A(\in \Gamma)$, representing a group of $t$ participants, is an authorized set.

The key components are distributed to the participants by a *receiver*. Then one group $A(\in \Gamma)$ of participants collaborate to perform a threshold operation such as encryption or authentication, using a key at their disposal, and sends the output of the shared operation back to the receiver. Once the receiver receives this message, he knows which key was used in the threshold operation. The objective is to prevent an adversary from learning which group of participants collaborated to perform the operation. The adversary may be the receiver or anyone knowing how key components are assigned to participants. So, given that a particular key was used, the issue of anonymity of the group $A(\in \Gamma)$ of participants which has performed the threshold operation is of importance. Similarly, the anonymity of any individual participant, as a member of that group, is also of concern. It is assumed that the communication of the message to the receiver is done using an anonymous channel and all details of communications among the participants remain unknown to the adversary.

Martin et al. (2003) gave a measure for group anonymity which essentially measures the average degree of anonymity for a scheme. In a recent work, Zaverucha and Stinson (2010; hereafter referred to as ZS) proposed an elegant new scheme for sharing symmetric key operations amongst a set of participants according to a $(t,n)$ threshold access structure. Focusing attention on access structures based on perfect hash families (PHFs), they explored at length the question of anonymity. They pointed out that anonymity should preferably be evaluated in the worst case instead of the average case and from this perspective, they gave a measure for group anonymity for their scheme, together with a measure of anonymity of any individual participant as a member of that group. They showed that their scheme ensures a higher level of anonymity than the existing ones. Accordingly, we consider this scheme of ZS as our starting point.

In this article, we adopt the measures of ZS and examine how, for a $(t,n)$ threshold access structure determined by any given PHF, further improvement in anonymity can be achieved. In Section 4, we propose a new scheme where, instead of all $A(\in \Gamma)$ being equally likely to use a key, as stipulated in the scheme of ZS, these groups $A$ now act with appropriately assigned unequal probabilities. We demonstrate that this new scheme has improved anonymity compared to the scheme of ZS and prove that it can, in fact, be optimal and equitable with respect to participant anonymity when based on a balanced PHF. In the process, in Section 3, we note that some results in ZS on participant anonymity require re-evaluation and we set them in the proper perspective. Finally, in Section 6, we show that our new scheme turns out to be quite promising even for threshold access structures which are given by combinatorial entities other than PHFs.

**2. Notation and Preliminaries**

For ease in reference, we begin with a brief description of PHFs and the associated $(t,n)$ threshold access structures; for details on sharing symmetric key operations we refer to ZS and the references therein. *A* PHF$(l;n,m,t)$, $m \geq t \geq 2$, *can be depicted as an* $l \times n$ *array, say B, populated with m symbols, say* $1,\ldots, m$, *such that within every* $l \times t$ *subarray of B, there is a row containing t distinct symbols*. Such a PHF is called *balanced* and is denoted by BPHF$(l;n,m,t)$ if the *m* symbols occur equally often in each row of *B*.

**Example 1**. It can be easily checked that the following $3 \times 6$ array represents a BPHF(3;6,2,2).

$$B = \begin{matrix} 1 & 1 & 1 & 2 & 2 & 2 \\ 1 & 1 & 2 & 1 & 2 & 2 \\ 1 & 2 & 2 & 1 & 1 & 2 \end{matrix}$$

Following ZS, this BPHF(3;6,2,2) leads to a scheme with a (2,6) threshold access structure having
(a) six [= (3)(2)] key components

$$(1,1), (1,2), (2,1), (2,2), (3,1), (3,2),$$

such that each row index of *B*, coupled with a symbol in *B*, gives a key component;

(b) three keys

$$K(1\times 12)=\{(1,1), (1,2)\}, \quad K(2\times 12)=\{(2,1), (2,2)\}, \quad K(3\times 12)=\{(3,1), (3,2)\},$$

such that each key is given by the Cartesian product of a row index of $B$ with the set of two symbols in that row of $B$, and

(c) six participants $P_1,...,P_6$, such that the key components given to any participant are dictated by the corresponding column of $B$, i.e., $P_1$ gets key components (1,1), (2,1), (3,1); $P_2$ gets key components (1,1), (2,1), (3,2); $P_3$ gets key components (1,1), (2,2), (3,2), and so on.

Thus, the pair of participants $P_1$ and $P_2$ can recover only the single key $K(3\times 12)$ while the pair $P_1$ and $P_3$ can recover both keys $K(2\times 12)$ and $K(3\times 12)$. Clearly, any two participants together hold at least one key while no single participant can recover any key. □

In general, a $\text{PHF}(l;n,m,t)$, when written as an $l\times n$ array $B$, leads to a scheme based on a $(t,n)$ threshold access structure involving

(a) $lm$ key components $(r, j)$, $1\le r\le l$, $1\le j\le m$;

(b) $l\binom{m}{t}$ keys denoted by $K(r\times J)$, say, $1\le r\le l$, $J\in\Delta$, where $\Delta$ denotes the collection of the $\binom{m}{t}$ $t$-tuples $j_1..j_t$ satisfying $1\le j_1<...<j_t\le m$, and for any $J=j_1...j_t\ (\in\Delta)$, the key $K(r\times J)$ consists of the $t$ key components $(r, j_1),...,(r, j_t)$;

(c) $n$ participants $P_1,...,P_n$ such that, for $1\le c\le n$, participant $P_c$ gets the $l$ key components $(r, b_{rc})$, $1\le r\le l$, where $(b_{1c},...,b_{rc})^T$ is the $c$-th column of $B$.

By (c), there is a one-to-one correspondence between the participants and the columns of $B$, and hence by (b), any group of $w$ participants can recover a specific key $K(r\times J)$, $J=j_1j_2...j_t$, *if and only if* the symbols $j_1, j_2,..., j_t$ appear in the $r$th row of the corresponding $l\times w$ subarray of $B$. Therefore, as noted in ZS, the following are evident:

- Every group $A\,(\in\Gamma)$ of participants can recover at least one key.
- No set of participants, involving less than $t$ members, can recover any key.

Let $s_A\ (\ge 1)$ denote the number of keys that the participants in $A\ (\in\Gamma)$ can collectively recover. In Example 1, clearly, $s_A$=1 for $A=\{P_1,P_2\}$, $s_A$=2 for $A=\{P_1,P_3\}$, and so on.

We now present the measures of group and participant anonymity as given in ZS. In connection with the use of any specific key, these quantify the level of confidentiality of the group of $t$ participants that have collaborated to use the key, or that of any individual participant as a member of this group. To that effect, for a scheme $S$, given that the key $K(r\times J)$ has been used, let $\Pr[A\,|\,K(r\times J)]$ denote the conditional probability that the group $A\ (\in\Gamma)$ of participants has used this key, and let $\Pr[P_c \mid K(r \times J)]$ denote the conditional probability that participant $P_c$ ($1 \leq c \leq n$) is in the group of participants that has used this key. Then the measures of group and participant anonymity of the scheme $S$ are given by

$$\mu = 1 - \max\{\Pr[A \mid K(r \times J)]: A \in \Gamma, 1 \leq r \leq l, J \in \Delta\}, \quad (1)$$

and

$$\rho = 1 - \max\{\Pr[P_c \mid K(r \times J)]: 1 \leq c \leq n, 1 \leq r \leq l, J \in \Delta\}, \quad (2)$$

respectively. In the same spirit, the anonymity of any particular participant $P_c$ is measured by

$$\rho(P_c) = 1 - \max\{\Pr[P_c \mid K(r \times J)]: 1 \leq r \leq l, J \in \Delta\}. \quad (3)$$

All these measures are intended to protect against the worst possible scenario. A useful scheme should aim at achieving larger values of the quantities in (1)-(3). Moreover, it is desirable that participant anonymity should be equitable, i.e., the values of $\rho(P_c), 1 \leq c \leq n$, should not differ significantly. Ideally, one should have $\rho(P_c) = 1 - (t/n), 1 \leq c \leq n$. A scheme achieving this is said to have optimal participant anonymity; cf. ZS.

## 3. ZS scheme revisited

ZS gave a scheme, focusing largely on BPHFs, and called this BPHF-MAC. They showed that this BPHF-MAC leads to improved anonymity compared to existing schemes. Their scheme is applicable also to constructions from PHFs which need not be balanced. We call their procedure the ZS scheme in general. In Section 4, we will modify the ZS scheme to ensure further gains in anonymity. Towards this, in this section, we revisit the ZS scheme in order to get a clear insight into its anonymity properties. In the process, a few examples are worked out in detail in order to indicate an anomaly in their results on participant anonymity.

With reference to a $(t,n)$ threshold access structure given by a $\text{PHF}(l;n,m,t)$, the ZS scheme has the following features:

(i) The $\binom{n}{t}$ groups of participants $A \ (\in \Gamma)$ are all equally likely to use a key.

(ii) If a group $A \ (\in \Gamma)$ uses a key, then it employs any one of its available $s_A$ keys with equal probability.

The main innovation by ZS lies in (ii) since in earlier schemes, any group $A$, with access to multiple keys of the form $K(r \times J)$, had to always use the key with smallest $r$. ZS showed that a modification as in (ii) entails gains in anonymity compared to the earlier schemes. We will show in Section 4 that a modification to (i) above leads to further gains in anonymity.

For any key $K(r \times J)$, let $Q(r \times J)$ denote the collection of the groups of participants that can recover this key, i.e.,

$$Q(r \times J) = \{A: A \in \Gamma, \text{ the participants in } A \text{ can recover the key } K(r \times J)\}. \quad (4)$$

Clearly, by (i) above, the unconditional probability for any group $A\,(\in \Gamma)$ to use a key is

$$\Pr[A] = 1/\binom{n}{t}, \qquad (5)$$

while by (ii), the conditional probability for the use of the key $K(r \times J)$, given that the group $A$ has used a key, is

$$\Pr[K(r \times J) \mid A] = s_A^{-1}, \text{ if } A \in Q(r \times J),$$
$$= 0, \quad \text{otherwise}, \qquad (6)$$

Hence, as also noted in equation (9) of ZS, in their scheme the key $K(r \times J)$ is used with unconditional probability

$$\Pr[K(r \times J)] = \frac{1}{\binom{n}{t}} \sum_{A \in Q(r \times J)} s_A^{-1} = \frac{1}{\binom{n}{t}} \sum_{G \in Q(r \times J)} s_G^{-1}. \qquad (7)$$

Therefore, using the standard conditional probability formula

$$\Pr[A \mid K(r \times J)] = \frac{\Pr[K(r \times J) \mid A] \Pr[A]}{\Pr[K(r \times J)]}, \qquad (8)$$

for any $A \in \Gamma$, under the ZS scheme, by (5)-(7) it follows that

$$\Pr[A \mid K(r \times J)] = \frac{s_A^{-1}}{\sum_{G \in Q(r \times J)} s_G^{-1}}, \text{ if } A \in Q(r \times J), \qquad (9)$$
$$= 0, \qquad \text{otherwise}.$$

This is essentially same as their equation (10). Furthermore, given that the key $K(r \times J)$ has been used, participant $P_c$ is in the group that has used this key if and only if $P_c \in A$ and $A$ has used the key, for some $A \in Q(r \times J)$. Thus for the ZS scheme,

$$\Pr[P_c \mid K(r \times J)] = \sum_{A \in Q(r \times J)} \delta_{cA} \Pr[A \mid K(r \times J)], \ 1 \leq c \leq n, \qquad (10)$$

where the indicator $\delta_{cA}$ equals 1 if $P_c \in A$, and 0 otherwise.

*3.1 Some Examples and their implications*

In their Section 3.2, ZS remarked that their BPHF-MAC scheme provides optimal and equitable participant anonymity and to establish this it was claimed in their Theorem 3.14 that for their scheme based on a BPHF$(l;n,t,t)$,

$$\Pr[P_c \mid K(r \times J)] = t/n, \text{ for every } c \text{ and every } r \times J, \qquad (11)$$

and that as a result

$$\rho(P_c) = 1 - (t/n), \text{ for every } c. \qquad (12)$$

Note that ZS had used the notation $\Pr[P_c \mid r]$ rather than $\Pr[P_c \mid K(r \times J)]$ in their Theorem 3.14 as $m$ equals $t$ in a PHF$(l;n,t,t)$, and hence $\Delta$ has a unique member $J = 12...t$. More generally, at the end of Section 4 of ZS, it was claimed that for their scheme based on a BPHF$(l;n,m,t)$

$$\rho(P_c) = 1 - (m/n), \text{ for every } c. \tag{13}$$

The next two examples demonstrate that the assertions (11)-(13) made in ZS about their scheme for BPHFs are not true in general.

**Example 1** (continued). This example is based on a BPHF(3;6,2,2). Consider first the key $K(1 \times 12)$. Note that $Q(1 \times 12)$ consists of nine pairs of participants, namely, $A(c,d) = \{P_c, P_d\}$, where $c \in \{1, 2, 3\}$ and $d \in \{4, 5, 6\}$, and these pairs can also recover varying numbers of other keys in addition to $K(1 \times 12)$. The $s_A$ values for these pairs $A(c,d)$ are as shown below:

| Pair | $A(1,4)$ | $A(1,5)$ | $A(1,6)$ | $A(2,4)$ | $A(2,5)$ | $A(2,6)$ | $A(3,4)$ | $A(3,5)$ | $A(3,6)$ |
|---|---|---|---|---|---|---|---|---|---|
| $s_A$ | 1 | 2 | 3 | 2 | 3 | 2 | 3 | 2 | 1 |

Hence (9) yields

$$\Pr[A(c,d) | K(1 \times 12)] = \tfrac{1}{5} \text{ for } (c,d) = (1, 4), (3, 6);$$
$$= \tfrac{1}{10} \text{ for } (c,d) = (1, 5), (2, 4), (2,6), (3, 5);$$
$$= \tfrac{1}{15} \text{ for } (c,d) = (1, 6), (2, 5), (3, 4).$$

Since $P_1$ belongs to $A(1,4)$, $A(1,5)$ and $A(1,6)$, it now follows from (10) that $\Pr[P_1 | K(1 \times 12)] = \tfrac{1}{5} + \tfrac{1}{10} + \tfrac{1}{15} = \tfrac{11}{30}$. Similarly,

$$\Pr[P_c | K(1 \times 12)] = \tfrac{11}{30} \text{ for } c = 1, 3, 4, 6;$$
$$= \tfrac{4}{15} \text{ for } c = 2, 5. \tag{14}$$

Turning to the keys $K(2 \times 12)$ and $K(3 \times 12)$, it may be verified in a similar manner that

$$\max\{\Pr[A | K(2 \times 12)]: A \in \Gamma\} = \max\{\Pr[A | K(3 \times 12)]: A \in \Gamma\} = \tfrac{1}{5}.$$

Also, $\Pr[P_c | K(2 \times 12)] = \tfrac{11}{30}$ or $\tfrac{4}{15}$, for $c = 2,3,4,5$, or $c = 1,6$, respectively, while $\Pr[P_c | K(3 \times 12)] = \tfrac{11}{30}$ or $\tfrac{4}{15}$, for $c = 1,2,5,6$, or $c = 3,4$, respectively. Hence by (1)-(3), we get in this example the anonymity measures

$$\mu = \tfrac{4}{5}, \qquad \rho = \tfrac{19}{30}, \qquad \rho(P_c) = \tfrac{19}{30}, 1 \le c \le 6, \tag{15}$$

for the ZS scheme. Here $t = 2$, $n = 6$, and the right-hand sides of (11) and (12) equal 1/3 and 2/3 respectively, which are not in agreement with (14) and (15) obtained via explicit calculation. □

**Example 2.** Consider a BPHF(3;18,6,3) represented by the $3 \times 18$ array

$$B = \begin{matrix} 1\ 6\ 3\ 2\ 6\ 2\ 1\ 5\ 2\ 5\ 1\ 4\ 5\ 6\ 3\ 3\ 4\ 4 \\ 1\ 6\ 5\ 1\ 5\ 4\ 2\ 4\ 3\ 1\ 6\ 2\ 2\ 3\ 6\ 4\ 3\ 5 \\ 5\ 2\ 6\ 2\ 3\ 3\ 3\ 4\ 6\ 1\ 4\ 2\ 6\ 5\ 1\ 5\ 1\ 4 \end{matrix}$$

accessed from the PHF tables by Walker (2011). Here $t = 3$ and as in Example 1, this $B$ gives a (3,18) threshold access structure with 18 [=(3)(6)] key components, 60 [= $3\binom{6}{3}$] keys and 18 participants $P_1, \ldots, P_{18}$ corresponding to the columns of $B$. Consider the key $K(1 \times 123) = \{(1,1), (1,2),$ (1,3)} and note that $Q(1\times 123)$ consists of 27 triplets of participants, namely, $A(c,d,e) = \{P_c, P_d, P_e\}$, where $c \in \{1, 7, 11\}$, $d \in \{4, 6, 9\}$ and $e \in \{3, 15, 16\}$. The $s_A$ values for these triplets are as follows:

| Triplet | A(1,4,3) | A(1,4,15) | A(1,4,16) | A(1,6,3) | A(1,6,15) | A(1,6,16) | A(1,9,3) |
|---|---|---|---|---|---|---|---|
| $s_A$ | 2 | 2 | 1 | 3 | 3 | 1 | 2 |
| Triplet | A(1,9,15) | A(1,9,16) | A(7,4,3) | A(7,4,15) | A(7,4,16) | A(7,6,3) | A(7,6,15) |
| $s_A$ | 3 | 2 | 3 | 3 | 3 | 2 | 2 |
| Triplet | A(7,6,16) | A(7,9,3) | A(7,9,15) | A(7,9,16) | A(11,4,3) | A(11,4,15) | A(11,4,16) |
| $s_A$ | 1 | 2 | 3 | 3 | 3 | 2 | 3 |
| Triplet | A(11,6,3) | A(11,6,15) | A(11,6,16) | A(11,9,3) | A(11,9,15) | A(11,9,16) | |
| $s_A$ | 3 | 2 | 2 | 2 | 2 | 3 | |

Hence with

$\Gamma_1 = \{(1,4,16), (1,6,16), (7,6,16)\}$,

$\Gamma_2 = \{(1,4,3), (1,4,15), (1,9,3), (1,9,16), (7,6,3), (7,6,15), (7,9,3), (11,4,15),$
$(11,6,15), (11,6,16), (11,9,3), (11,9,15)\}$,

$\Gamma_3 = \{(1,6,3), (1,6,15), (1,9,15), (7,4,3), (7,4,15), (7,4,16), (7,9,15), (7,9,16),$
$(11,4,3), (11,4,16), (11,6,3), (11,9,16)\}$,

it follows from (9) that

$$\Pr[A(c,d,e) \mid K(1\times 123)] = \tfrac{1}{13} \text{ for } (c,d,e) \in \Gamma_1,$$
$$= \tfrac{1}{26} \text{ for } (c,d,e) \in \Gamma_2,$$
$$= \tfrac{1}{39} \text{ for } (c,d,e) \in \Gamma_3.$$

Since $P_1$ belongs to $A(1,d,e)$ for $d \in \{4, 6, 9\}$ and $e \in \{3, 15, 16\}$, now (10) yields

$$\Pr[P_1 \mid K(1\times 123)] = 2 \times \tfrac{1}{13} + 4 \times \tfrac{1}{26} + 3 \times \tfrac{1}{39} = \tfrac{30}{78}.$$

Similarly,

$$\Pr[P_c \mid 1\times 123] = \tfrac{32}{78} \text{ for } c = 16;$$
$$= \tfrac{30}{78} \text{ for } c = 1, 6;$$
$$= \tfrac{25}{78} \text{ for } c = 4, 7;$$
$$= \tfrac{23}{78} \text{ for } c = 3, 9, 11, 15;$$
$$= 0 \text{ for every other } c.$$

In view of the aforesaid expressions for $\Pr[A(c,d,e) \mid K(1\times 123)]$ and $\Pr[P_c \mid K(1\times 123)]$, even without considering the other keys, it follows from (1) and (2) that in this example

$$\mu \leq \tfrac{12}{13}, \quad \rho \leq \tfrac{46}{78}, \tag{16}$$

for the ZS scheme. Here $m = 6$, $n = 18$, and the right-hand side of (13) equals 2/3, implying $\rho = 2/3$, which is not in agreement with (16) obtained via explicit calculation. □

Examples 1 and 2 demonstrate a conflict between the reality and equations (11)-(13) as claimed in ZS. This can be attributed to the fact that the derivation of these equations in ZS does not account for the quantities $s_A$ which play a crucial role in their scheme in the sense that any group of participants $A (\in \Gamma)$ uses one of its $s_A$ keys with equal probability. In some situations, however, the values of the anonymity measures obtained by (11)-(13) may coincide with the values in reality, as is the case in Example 3 below.

**Example 3.** Consider a BPHF(4;9,3,3) represented by the $4 \times 9$ array (cf. ZS)

$$B = \begin{matrix} 1 & 3 & 2 & 2 & 3 & 2 & 3 & 1 & 1 \\ 1 & 3 & 1 & 3 & 1 & 2 & 2 & 2 & 3 \\ 1 & 2 & 2 & 1 & 3 & 3 & 1 & 2 & 3 \\ 3 & 3 & 2 & 1 & 1 & 3 & 2 & 1 & 2 \end{matrix}$$

which gives a (3, 9) threshold access structure with 12 key components, 4 keys and 9 participants $P_1,...,P_9$. Calculations similar to those in Examples 1 and 2 show that here $\mu = \frac{20}{21}$, $\rho = \frac{2}{3}$ and $\rho(P_c) = \frac{2}{3}, 1 \leq c \leq 9$, for the ZS scheme. □

While the $\rho(P_c)$ values in Example 3 are consistent with (12), this coincidence is only the result of some kind of symmetry inherent in the associated BPHF; such symmetry may not hold for all BPHFs in general, as illustrated in Examples 1 and 2. The points noted in this section underscore the need to set (11)-(13) in the proper perspective. Does there exist a scheme which will be truly optimal or equitable with respect to participant anonymity in the sense of ensuring the validity of (11)-(13) for a $(t,n)$ threshold access structure given by any BPHF? In Section 4, we will present such a new scheme which will also entail gains in anonymity, compared to the ZS scheme, for both balanced and unbalanced PHFs; cf. Remarks 2-4 in that section.

### 3.2. Some inequalities related to anonymity measures

We continue with the ZS scheme, as given by a PHF($l;n,m,t$) which need not necessarily be balanced, and present some inequalities on the anonymity properties of this scheme. These will be needed in Section 4 for comparative purposes. As before, let $B = (b_{rc})$, $1 \leq r \leq l$, $1 \leq c \leq n$, be the $l \times n$ array depicting the PHF. For $1 \leq r \leq l$ and $1 \leq j \leq m$, write

$$F(r,j) = \{c: 1 \leq c \leq n, b_{rc} = j\}, \tag{17}$$

i.e., $F(r,j)$ is the index set of participants $P_c$ who get the key component $(r,j)$. Let $f(r,j)$ be the cardinality of $F(r,j)$ and, to avoid trivialities, assume that $f(r,j) \geq 1$, for every $r$ and $j$. Write $q(r \times J)$ for the cardinality of the set $Q(r \times J)$ defined in (4), $1 \leq r \leq l$, $J \in \Delta$.

**Lemma 1.** *Consider the ZS scheme based on a* PHF($l;n,m,t$). *Then for every* $1 \leq r \leq l$ *and* $J = j_1...j_t$ $(\in \Delta)$,

(a) $\max\{\Pr[A\,|\,K(r\times J)]:\ A\in\Gamma\}\geq\{q(r\times J)\}^{-1}$.

(b) $\max\{\Pr[P_c\,|\,K(r\times J)]:c\in F(r,j_i)\}\geq\{f(r,j_i)\}^{-1}$, $1\leq i\leq t$.

*Proof.* (a) By (9), $\Pr[A\,|\,K(r\times J)]=0$, for $A\notin Q(r\times J)$, so that

$$\max\{\Pr[A\,|\,K(r\times J)]:\ A\in\Gamma\}=\max\{\Pr[A\,|\,K(r\times J)]:\ A\in Q(r\times J)\}. \qquad (18)$$

Now by (9), the sum of the $q(r\times J)$ quantities $\Pr[A\,|\,K(r\times J)]$, over $A\in Q(r\times J)$, equals 1, i.e., their arithmetic mean equals $\{q(r\times J)\}^{-1}$. Hence (a) follows from (18).

(b) Consider any fixed $i$ $(1\leq i\leq t)$. Given that the key $K(r\times J)$, $J=j_1...j_t$, has been used, exactly one member of the group of participants using the key must be having the key component $(r,j_i)$, i.e., by (17), one and only one participant $P_c$, with $c\in F(r,j_i)$, certainly belongs to this group. Therefore, the sum of the $f(r,j_i)$ quantities $\Pr[P_c\,|\,K(r\times J)]$, over $c\in F(r,j_i)$, equals 1. Hence (b) follows arguing as in (a). □

From Lemma 1 and the definitions of $\mu$ and $\rho$ in (1) and (2), the following is evident.

**Lemma 2.** *For the ZS scheme based on a* PHF$(l;n,m,t)$,

(a) $\mu\leq 1-[\min\{q(r\times J):1\leq r\leq l,\ J\in\Delta\}]^{-1}$.

(b) $\rho\leq 1-[\min\{f(r,j):1\leq r\leq l,1\leq j\leq m\}]^{-1}$. □

**4. A new scheme: the proportional scheme**

With reference to a $(t,n)$ threshold access structure given by a PHF$(l;n,m,t)$ which may or may not be balanced, we propose the following new scheme:

(i)′ Any group of participants $A\ (\in\Gamma)$ uses a key with probability proportional to $s_A$.

(ii) If a group $A\ (\in\Gamma)$ uses a key, then it employs any of its available $s_A$ keys with equal probability.

While (ii) is the same as that for the ZS scheme, (i)′ differs from their (i) and takes cognizance of the variation in the $s_A$ values across $A\in\Gamma$. This modification will be seen to have useful consequences; in particular, when based on a BPHF$(l;n,m,t)$, this new scheme provides completely equitable participant as well as group anonymity. Moreover, when $m=t$, this scheme ensures optimal participant anonymity too.

Hereafter, this new scheme given by (i)′ and (ii) will be referred to as the *proportional* scheme. An algorithm for its easy implementation is presented in the next section. We now examine the behavior of this proportional scheme with respect to anonymity.

**Theorem 1**. *Consider the proportional scheme based on a* $\text{PHF}(l; n, m, t)$. *Then for every* $1 \leq r \leq l$ *and* $J = j_1 \ldots j_t$ $(\in \Delta)$,

(a) $\Pr[A \mid K(r \times J)] = \{q(r \times J)\}^{-1}$, *if* $A \in Q(r \times J)$,
$\qquad \qquad \quad = 0,$ *otherwise.*

(b) $\Pr[P_c \mid K(r \times J)] = \{f(r, j_i)\}^{-1}$, *if* $c \in F(r, j_i)$, $1 \leq i \leq t$,
$\qquad \qquad \quad = 0,$ *otherwise.*

*Proof.* (a) Using the same notation as in Section 3, by (i)',

$$\Pr[A] = s_A / s_0, \qquad (19)$$

where $s_0$ equals the sum of the quantities $s_A$, over $A \in \Gamma$. Also, (6) continues to hold because of (ii). Therefore, recalling that $q(r \times J)$ is the cardinality of $Q(r \times J)$, we get

$$\Pr[K(r \times J)] = \sum_{A \in Q(r \times J)} (s_A / s_0) s_A^{-1} = \frac{q(r \times J)}{s_0}, \qquad (20)$$

as a counterpart of (7). Now (a) follows from (6), (19) and (20) invoking the conditional probability formula (8).

(b) In view of (a), for $1 \leq c \leq n$, as a counterpart of (10), we get

$$\Pr[P_c \mid K(r \times J)] = \sum_{A \in Q(r \times J)} \delta_{cA} \Pr[A \mid r \times J] = \frac{1}{q(r \times J)} \sum_{A \in Q(r \times J)} \delta_{cA}. \qquad (21)$$

By (4), any group $A$ belongs to $Q(r \times J)$, $J = j_1 \ldots j_t$, if and only if it includes exactly one participant having the key component $(r, j_i)$, $1 \leq i \leq t$. In view of (17), therefore, $Q(r \times J)$ consists of the groups of participants

$$A(c(1), c(2), \ldots, c(t)) = \{P_{c(1)}, P_{c(2)}, \ldots, P_{c(t)}\}, \quad c(i) \in F(r, j_i), 1 \leq i \leq t, \qquad (22)$$

and hence

$$q(r \times J) = \prod_{i=1}^{t} f(r, j_i). \qquad (23)$$

Consider now a particular participant $P_c$, with $c \in F(r, j_1)$. By (22), there are $\prod_{i=2}^{t} f(r, j_i)$ groups of participants, namely $A(c, c(2), \ldots, c(t))$, $c(i) \in F(r, j_i), 2 \leq i \leq t$, which include $P_c$ and belong to $Q(r \times J)$. Thus for such a $P_c$

$$\sum_{A \in Q(r \times J)} \delta_{cA} = \prod_{i=2}^{t} f(r, j_i),$$

so that by (21) and (23), $\Pr[P_c \mid K(r \times J)] = \{f(r, j_1)\}^{-1}$. Similarly, the conclusion of (b) holds for any $P_c$, with $c \in F(r, j_i)$, $2 \leq i \leq t$. Finally, if $P_c$ is such that $c \notin F(r, j_i)$ for every $i$ $(1 \leq i \leq t)$, then none of the groups, constituting $Q(r \times J)$ and listed in (22), can include $P_c$. As a result, then $\Pr[P_c | K(r \times J)] = 0$ by (21). This completes the proof of (b). □

The next result is immediate from Theorem 1 and the definitions of $\mu$ and $\rho$ in (1) and (2).

**Theorem 2.** *For the proportional scheme based on a* $\text{PHF}(l; n, m, t)$,

(a) $\mu = 1 - [\min\{q(r \times J) : 1 \leq r \leq l, J \in \Delta\}]^{-1}$.

(b) $\rho = 1 - [\min\{f(r, j) : 1 \leq r \leq l, 1 \leq j \leq m\}]^{-1}$. □

**Remark 1**. By (17), $\Sigma_{j=1}^{m} f(r, j) = n$, for $1 \leq r \leq l$. This, in conjunction with (23), implies that

$$\min\{f(r, j) : 1 \leq r \leq l, 1 \leq j \leq m\} \leq n/m,$$

and $\min\{q(r \times J) : 1 \leq r \leq l, J \in \Delta\} \leq (n/m)^t$,

equality being attained in both cases *if and only if* $f(r, j) = n/m$ for every $r$ and $j$, in which case the PHF is balanced. Since $m \geq t$, it follows that the expressions for $\mu$ and $\rho$ in Theorem 2 satisfy

$$\mu \leq 1 - (m/n)^t \leq 1 - (t/n)^t \quad \text{and} \quad \rho \leq 1 - (m/n) \leq 1 - (t/n).$$

Therefore, from consideration of anonymity, given *l, n* and *t*, the proportional scheme should ideally be based on a $\text{BPHF}(l; n, t, t)$ if it exists; otherwise, one should attempt to keep *m* close to *t* and the $f(r, j)$ as nearly equal as possible. □

From (3) and Theorems 1 and 2, we get Corollary 1 below summarizing, in continuation of Remark 1, the anonymity properties of the proportional scheme, when based on a $\text{BPHF}(l; n, m, t)$.

**Corollary 1**. *Consider the proportional scheme based on a* $\text{BPHF}(l; n, m, t)$. *Then*

(a) *For every* $1 \leq r \leq l$ *and* $J = j_1 ... j_t \ (\in \Delta)$,

(i) $\Pr[A | K(r \times J)] = (m/n)^t$, if $A \in Q(r \times J)$,
$\qquad\qquad\qquad\qquad = 0, \quad$ *otherwise*.

(ii) $\Pr[P_c | K(r \times J)] = m/n$, if $c \in F(r, j_i)$, $1 \leq i \leq t$,
$\qquad\qquad\qquad\qquad = 0, \quad$ *otherwise*.

(b) $\mu = 1 - (m/n)^t$.
(c) $\rho = 1 - (m/n)$.
(d) $\rho(P_c) = 1 - (m/n)$, $1 \leq c \leq n$. □

The above results have major implications as remarked below.

**Remark 2:** Lemma 2 and Theorem 2 show that, for any arbitrary $\text{PHF}(l; n, m, t)$, balanced or not, the values of the anonymity measures $\mu$ and $\rho$ for the proportional scheme are never less than their counterparts under the ZS scheme. Indeed, as will be seen later through examples, the proportional scheme often yields larger values of $\mu$ and $\rho$, thus leading to gains in anonymity. □

**Remark 3:** Corollary 1(d) shows that for any $\text{BPHF}(l; n, m, t)$, the proportional scheme ensures the attainment of (13) and hence provides *completely equitable participant anonymity*. Furthermore, when $m = t$, by Corollary 1(a)(ii) and (d), this scheme ensures the attainment of (11) and (12) as well and hence provides *optimal participant anonymity* too. In this connection, note that if $m = t$, then $\Delta$ has a unique member $J = 12...t$, so that in this case, the union of the sets $F(r, j_i)$, $1 \leq i \leq t$, equals $\{1, 2,..., n\}$ and Corollary 1(a)(ii) reduces to (11). Thus, in addition to enhancing anonymity, the proportional scheme also guarantees the assertions in (11)-(13) which, as seen earlier, do not hold in general for the ZS scheme based on BPHFs. □

**Remark 4:** Corollary 1(a)(i) shows that for a $BPHF(l; n, m, t)$, the proportional scheme ensures *completely equitable group anonymity* as well, in the sense of achieving

$$\max\{\Pr[A \mid K(r \times J)]: 1 \leq r \leq l, J \in \Delta\} = (m/n)^t, \text{ for every } A (\in \Gamma).$$

This property is not shared by the ZS scheme – e.g., in Example 1, given by a $BPHF(3;6,2,2)$, it may be checked that the quantities $\max\{\Pr[A \mid K(r \times J)]: 1 \leq r \leq l, J \in \Delta\}$ can equal $\frac{1}{5}, \frac{1}{10}$ or $\frac{1}{15}$, depending on the choice of $A (\in \Gamma)$, under the ZS scheme. □

In order to illustrate the point noted in Remark 2 above, we give a few examples. For this, we first go back to Examples 1-3 and compare $\mu$ and $\rho$ for the ZS and the proportional schemes. These examples are based on BPHFs and hence Corollary 1 is applicable to the latter scheme.

| Example | ZS scheme | Proportional scheme |
|---|---|---|
| Example 1 | $\mu = \frac{4}{5}$, $\rho = \frac{19}{30}$ | $\mu = \frac{8}{9}$, $\rho = \frac{2}{3}$ |
| Example 2 | $\mu \leq \frac{12}{13}$, $\rho \leq \frac{23}{39}$ | $\mu = \frac{26}{27}$, $\rho = \frac{2}{3}$ |
| Example 3 | $\mu = \frac{20}{21}$, $\rho = \frac{2}{3}$ | $\mu = \frac{26}{27}$, $\rho = \frac{2}{3}$ |

Clearly, the $\mu$ for the proportional scheme is larger than that for the ZS scheme in all these examples. The same happens with $\rho$ as well, except in Example 3 where there is a tie between the two.

The above examples are for balanced PHFs. We now give two examples to compare these schemes for unbalanced PHFs.

**Example 4.** Consider a $PHF(4;9,2,2)$ represented by the $4 \times 9$ array

$$B = \begin{matrix} 1 & 1 & 1 & 1 & 2 & 2 & 2 & 2 & 2, \\ 1 & 1 & 1 & 2 & 1 & 1 & 2 & 2 & 2 \\ 1 & 1 & 2 & 2 & 1 & 1 & 1 & 2 & 2 \\ 1 & 2 & 2 & 2 & 2 & 1 & 1 & 1 & 2 \end{matrix}$$

which gives a (2, 9) threshold access structure with 8 key components, 4 keys and 9 participants. Here the pair $(f(r,1), f(r,2))$ equals (4, 5) or (5, 4) for every $r$. Hence by (23) and Theorem 2, for the proportional scheme, $\mu = \frac{19}{20}$ and $\rho = \frac{3}{4}$. Both these are larger than the corresponding quantities $\mu = \frac{23}{26}$ and $\rho = \frac{73}{104}$ for the ZS scheme, as calculations similar to those in Example 1 show. □

**Example 5.** Consider a $PHF(3;12,5,3)$ represented by the $3 \times 12$ array (cf. ZS)

$$B = 2\ 4\ 4\ 4\ 5\ 1\ 5\ 2\ 3\ 3\ 3\ 1,$$
$$3\ 1\ 5\ 4\ 5\ 1\ 2\ 5\ 2\ 4\ 3\ 3$$
$$1\ 4\ 5\ 1\ 3\ 2\ 1\ 2\ 4\ 2\ 3\ 5$$

which gives a (3,12) threshold access structure with 15 key components, 30 keys and 12 participants $P_1,\ldots,P_{12}$. For the ZS scheme, calculations similar to those in Example 2 yield

$$\max\{\Pr[A\,|\,2\times 124]: A\in\Gamma\} = \tfrac{3}{17} \quad \text{and} \quad \max\{\Pr[P_c\,|\,2\times 124]: 1\le c\le 12\}=\tfrac{10}{17},$$

so that by (1) and (2), $\mu \le \tfrac{14}{17}$ and $\rho \le \tfrac{7}{17}$. Here the vector $(f(r,1),\ldots,f(r,5))$ equals $(2, 2, 2, 3, 3)$ or a permutation thereof for every $r$. Hence by (23) and Theorem 2, for the proportional scheme, $\mu = \tfrac{7}{8}$ and $\rho = \tfrac{1}{2}$. Again, both $\mu$ and $\rho$ are larger than what the ZS scheme achieves. □

## 5. Algorithmic implementation of the proportional scheme

We now present a simple algorithm for the implementation of the proportional scheme based on a PHF$(l;n,m,t)$ as depicted by an $l\times n$ array $B$. The $l\times t$ subarray of $B$ corresponding to any group $A\,(\in\Gamma)$ of $t$ participants will be denoted by $B(A)$. Any row of $B(A)$ having $t$ distinct symbols is said to separate $A$ and determines a key that $A$ can recover – e.g., if the $r$th row of $B(A)$ has distinct symbols $j_1,\ldots,j_t$ $(j_1 < \ldots < j_t)$, then $A$ can recover the key $K(r\times J)$, $J = j_1\ldots j_t$.

*Algorithm*

I. One of the $\binom{n}{t}$ groups in $\Gamma$, say $A$, is activated at random with equal probability.

II. Group $A$ selects one of the rows of $B(A)$ at random with equal probability. If this row separates $A$, then group $A$ goes to Step III; else, the algorithm returns to Step I.

III. Group $A$ employs the key as determined by the separating row in Step II.

This algorithm is very similar to the one for the ZS scheme (ZS, p. 148) with the *only* change being that here selection of a non-separating row in Step II returns the algorithm to Step I, whereas in ZS selection of a non-separating row in Step II returns the algorithm to the beginning of Step II itself. The next result establishes that our algorithm, indeed, implements the proportional scheme.

**Theorem 3**. *The above algorithm ensures the following:*

(a) *Any group of participants $A\,(\in\Gamma)$ uses a key with probability proportional to $s_A$.*

(b) *If a group $A\,(\in\Gamma)$ uses a key, then it employs any one of its available $s_A$ keys with equal probability.*

*Proof.* For notational simplicity, let $N = \binom{n}{t}$. Then the probability that any one cycle of Steps I and II ends with group $A$ going to Step III is equal to $N^{-1}s_A/l$. Similarly, the probability that such a cycle is inconclusive (i.e., ends with a return to Step I) is

$$\sum_{G\in\Gamma} N^{-1}\{1-(s_G/l)\} = 1-(N^{-1}s_0/l),$$

where $s_0$ is the sum of the quantities $s_G$, over $G \in \Gamma$. Now, group $A$ eventually makes use of a key if and only if there is one cycle of Steps I and II that ends with group $A$ going to Step III, all previous cycles being inconclusive. Hence the probability for group $A$ to use a key is

$$\sum_{j=1}^{\infty} \{1-(N^{-1}s_0/l)\}^{j-1}(N^{-1}s_A/l) = s_A/s_0,$$

which proves (a). Part (b) is evident from symmetry considerations. □

## 6. Beyond PHFs

The proportional scheme appears to be quite promising even for $(t,n)$ threshold access structures which are given by combinatorial entities other than PHFs. Consider a general setup with

(i) $p$ key components 1, 2, …, $p$,

(ii) $v$ keys $K_1,...,K_v$, each of which is a set in $\{1, 2,..., p\}$ and which together cover $\{1, 2,..., p\}$, and

(iii) $n$ participants $P_1,...,P_n$ among whom the key components are distributed,

such that every group of $t$ participants can recover at least one key, and no set of participants, involving less than $t$ members, can recover any key.

As before, let $\Gamma$ be the collection all groups of $t$ participants. Analogously to (1)-(3), we now consider

$$\mu = 1 - \max\{\Pr[A \mid K_i]: A \in \Gamma, 1 \leq i \leq v\}, \tag{24}$$

$$\rho = 1 - \max\{\Pr[P_c \mid K_i]: 1 \leq c \leq n, 1 \leq i \leq v\}, \tag{25}$$

$$\rho(P_c) = 1 - \max\{\Pr[P_c \mid K_i]: 1 \leq i \leq v\}, \tag{26}$$

as measures of overall group anonymity, overall participant anonymity, and anonymity for participant $P_c$, respectively. Here the quantities $\Pr[A \mid K_i]$ and $\Pr[P_c \mid K_i]$ are obvious counterparts of $\Pr[A \mid K(r \times J)]$ and $\Pr[P_c \mid K(r \times J)]$ as defined earlier. Let $q_i$ denote the number of groups of $t$ participants that can recover the key $K_i$, $1 \leq i \leq v$.

The same arguments as those leading to Lemma 2(a) and Theorem 2(a) now show that

$$\mu \leq 1 - [\min\{q_i: 1 \leq i \leq v\}]^{-1},$$

for the ZS scheme, whereas

$$\mu = 1 - [\min\{q_i: 1 \leq i \leq v\}]^{-1}, \tag{27}$$

for the proportional scheme. Thus the value of $\mu$ for the proportional scheme is never less than that under the ZS scheme. However, no such statement can be made in full generality about $\rho$ because counterparts of the sets $F(r, j)$ in (17) are no longer available here, thus precluding any extension of Lemma 2(b) and Theorem 2(b). Nevertheless, even in the general setup, the proportional scheme can perform better than the ZS scheme in respect of both $\mu$ and $\rho$. An illustrative example follows.

**Example 6.** Let $t = 3$ and $p = n = 7$. The key components distributed to participant $P_c, 1 \leq c \leq 7$, are as listed in the $c$th column of the array

$$P = \begin{matrix} 1 & 2 & 3 & 4 & 5 & 6 & 7, \\ 2 & 3 & 4 & 5 & 6 & 7 & 1 \\ 4 & 5 & 6 & 7 & 1 & 2 & 3 \end{matrix}$$

i.e., $P_1$ gets key components 1, 2, 4, while $P_2$ gets 2, 3, 5, and so on. There are $v = 7$ keys $K_i$, $1 \leq i \leq 7$, where $K_i$ includes all key components except $i$. Thus each $K_i$ contains six key components. The columns of $P$, when interpreted as blocks, form a balanced incomplete block design. Since any two distinct participants have exactly one common key component, it is easy to verify that the above gives a (3,7) threshold access structure, indeed. As in Example 2, let $A(c,d,e) = \{P_c, P_d, P_e\}$. Then the keys that various triplets of participants can recover are as shown below.

| Triplets | Keys that can be recovered |
|---|---|
| $A(2,3,4), A(2,3,6), A(2,4,6), A(3,4,6)$ | Only $K_1$ |
| $A(3,4,5), A(3,4,7), A(3,5,7), A(4,5,7)$ | Only $K_2$ |
| $A(1,4,5), A(1,4,6), A(1,5,6), A(4,5,6)$ | Only $K_3$ |
| $A(2,5,6), A(2,5,7), A(2,6,7), A(5,6,7)$ | Only $K_4$ |
| $A(1,3,6), A(1,3,7), A(1,6,7), A(3,6,7)$ | Only $K_5$ |
| $A(1,2,4), A(1,2,7), A(1,4,7), A(2,4,7)$ | Only $K_6$ |
| $A(1,2,3), A(1,2,5), A(1,3,5), A(2,3,5)$ | Only $K_7$ |
| $A(1,2,6), A(1,3,4), A(1,5,7), A(2,3,7), A(2,4,5), A(3,5,6), A(4,6,7)$ | All seven keys |

Consider first the ZS scheme and the key $K_1$. The above table exhibits the $q_1 = 11$ triplets that can recover $K_1$; of these, four can recover only $K_1$ and the rest can recover all the seven keys. Hence proceeding as in (9) and (10), we get $\max\{\Pr[A|K_1]: A \in \Gamma\} = \frac{1}{5}$, and

$$\Pr[P_c | K_1] = \tfrac{24}{35}, \text{ for } c = 2, 3, 4, 6;$$
$$= \tfrac{3}{35}, \text{ for } c = 1, 5, 7.$$

Based on similar calculations for the other keys as well, one can invoke (24)-(26) to check that $\mu = \tfrac{4}{5}$, $\rho = \tfrac{11}{35}$ and $\rho(P_c) = \tfrac{11}{35}, 1 \leq c \leq 7$, for the ZS scheme.

We next turn to the proportional scheme. Then (27) yields $\mu = \tfrac{10}{11}$, since $q_i = 11$ for every $i$. Furthermore, analogously to Theorem 1(a), $\Pr[A|K_i] = \tfrac{1}{11}$, for every $A \in Q_i$, where $Q_i$ represents the collection of triplets of participants that can recover $K_i$, $1 \leq i \leq 7$. Hence, invoking (25) and

(26) again, it is not hard to see from the above table that $\rho = \frac{5}{11}$ and $\rho(P_c) = \frac{5}{11}$, $1 \leq c \leq 7$, for the proportional scheme. Thus in this example, the proportional scheme makes the values of $\mu$, $\rho$ and each $\rho(P_c)$ much larger than those under the ZS scheme. □

The last example highlights the need for further exploration of the behavior of the proportional scheme vis-à-vis the ZS scheme with reference to the general setup of this section. While this is beyond the scope of the present article which is focused on schemes based on PHFs, we conclude with the hope that our endeavor will generate interest in this and related issues.

**Acknowledgement.** The work of Rahul Mukerjee was supported by the J.C. Bose National Fellowship of the Government of India and a grant from the Indian Institute of Management Calcutta.